\documentstyle[12pt]{article}
\def\pa{\partial}
\title{Covariant Supplementation Scheme for Infinitely
Reducible First Class Constraints}
\author{A.A. Deriglazov\thanks{E-mail:
deriglaz@phys.tsu.tomsk.su}  and A.V. Galajinsky}
\date{Department of Theoretical Physics, Tomsk State University,
634050 Tomsk, Russia}
\begin{document}
\maketitle

\begin{abstract}
For a rather broad class of dynamical systems subject to mixed
fermionic first and second class constraints or infinitely reducible
first class constraints (IR1C), a manifestly covariant scheme of
supplementation of IR1C to irreducible ones is proposed. For a model
with IR1C only, an application of the scheme leads to a system with
covariantly splitted and irreducible first and second class
constraints. Modified Lagrangian formulations for the Green--Schwarz
superstring, Casalbuoni--Brink--Schwarz superparticle and Siegel
superparticle, which reproduce the supplementation scheme, are
suggested.
\end{abstract}


The problem of constructing a covariant quantization scheme for dynamical
systems with mixed first and second class constraints is extremely
urgent since the Green--Schwarz (GS) superstring [1] and
Casalbuoni--Brink--Schwarz (CBS) superparticle [2] belong to this class of
theories.\footnote{We mainly discuss the case of $D=10, N=1$ superspace
for which there is no a Poincar\'e covariant and irreducible splitting of the original
fermionic constraints on first and second class in the initial phase space.}
A general recipe of Hamiltonian quantization without
explicit splitting of the constraints has been
developed in a series of works by Batalin and Tyutin [3]. However, as it was
shown in the recent paper [4], an application of the scheme for
concrete models may conflict with manifest Poincar\'e covariance.

An alternative possibility for the theories concerned consists in
making use of covariant projectors to get splitted and reducible
subsets of first and second class constraints. Projectors with desired
properties have been constructed for the GS superstring [5], $D=9$
massive superparticle [6, 4], and $D=10,N=1$ CBS superparticle [7, 4].
This reduces the problem to quantization of linearly dependent second
class constraints (2CC) (which can be treated in covariant
fashion along the lines of Refs. 4, 7) and to quantization of infinitely
reducible first class constraints (IR1C). Unfortunately, the direct
application of BFV--BV methods in the latter case leads to the formulations
involving infinite extra ghost tower (see [8, 9] and references
therein), what extremely complicates the analysis of BRST cogomologies
and constructing effectively calculable quantum action.

In this letter, within the Hamiltonian and Lagrangian
framework, we suggest a covariant scheme of supplementation of
fermionic IR1C to a constraints system of finite stage of reducibility. In
Hamiltonian approach, the initial phase space is enlarged by auxiliary
variables, whose nondynamical character is provided by new reducible
constraints. The proposed trick is based on a
possibility to combine IR1C of extended formulation into covariant
first class constraints system of finite stage of
reducibility. After that, the standard quantization technique may be
employed [10], in particular, with a finite number of
ghost variables.

Two different cases will be considered: (i) models with mixed first and
second class constraints; (ii) models with IR1C only (the latter situation
takes place for modifications of the superstring and superparticle due
to Siegel [11, 12] and their generalizations [7, 13]). In the first case,
the resultant modified formulation contains irreducible first class
constraints (1CC) and separated from them linearly dependent 2CC. In
the second case, we shall get an extended system with irreducible 2CC and
splitted from them 1CC no more than of first stage of reducibility.
Although the presented scheme can be directly applied to constrained
systems of special form only (see Eq. (1) below), a class of these
theories is broad enough, in particular it includes all the above
mentioned superstring and superparticle models.

In our opinion, the advantages of the proposed trick consist in the
following: (i) Relatively small number of auxiliary variables are
needed as compared to the combined harmonic-twistor approaches
[14--16]. Note also that no twistor-like variables are introduced.
(ii) There exists a covariant gauge for both the initial and
auxiliary variables (compare with Refs. 14 and 17).

A consistent treatment of deformed constraints system implies the
construction of modified Lagrangian formulation on enlarged
configuration space which will reproduce the supplementation scheme.
The existence of such a formulation will allow, in particular, to prove
an equivalence of modified and
initial models. The corresponding Lagrangian formulations for the GS
superstring, CBS superparticle, and Siegel superparticle are presented
and analyzed in the letter.

We work in 16-component formalism of the Lorentz group $SO(1{,}9)$,
then $\theta^\alpha$, $\psi_\alpha$, $\alpha=1,\dots,16$, are
Majorana--Weyl spinors of opposite chirality. Real, symmetric
$16\times16$ $\Gamma$-matrices ${\Gamma^\mu}_{\alpha\beta}$,
$\tilde\Gamma^{\mu\alpha\beta}$ obeying the algebra
$\Gamma^\mu\tilde\Gamma^\nu+\Gamma^\nu\tilde\Gamma^\mu=-2\eta^{\mu\nu}$
will be used. Momenta conjugate to configuration space variables $c^i$
are denoted as $p_{ci}$.

Let us consider a dynamical system with fermionic pairs $(\theta^\alpha,
p_{\theta\alpha})$ being presented among the phase space variables $z^A$.
It is supposed that a complete constraints system of the model includes
(among others) the following:
\begin{equation}
L_\alpha\equiv p_{\theta\alpha}- iB_\mu{\Gamma^\mu}_{\alpha\beta}
\theta^\beta\approx 0, \qquad D^\mu D_\mu\approx0,
\end{equation}
and the Poisson bracket of the fermionic constraints is
\begin{equation}
\{L_\alpha,L_\beta\}=2iD_\mu{\Gamma^\mu}_{\alpha\beta}.
\end{equation}
Here, the $B^\mu(z)$ and $D^\mu(z)$ are some functions of phase variables so
that $D^2\approx0$ is first class constraint. The full set of
constraints of the theory may include ones different from Eq. (1), which
are inessential for subsequent analysis. Note that Eqs. (1), (2)
correspond to the CBS superparticle if we choose $B^\mu=D^\mu=p^\mu$,
where $p^\mu$ are momenta conjugated to space-time coordinates $x^\mu$.
To get the GS superstring case we choose
$B^\mu\equiv p^\mu+ \pa_1x^\mu-i\theta\Gamma^\mu\pa_1\theta$,
$D^\mu\equiv p^\mu+\pa_1x^\mu -2i\theta\Gamma^\mu\pa_1\theta$ (see
below). From Eqs. (1), (2) it follows that there are eight 1CC and
eight 2CC among the equations $L_\alpha\approx0$. To separate them in a
manifestly covariant fashion, let us extend the initial phase space by
a pair of vector variables $(\Lambda^\mu, p_{\Lambda\mu})$ subject to
constraints
\begin{equation}
\Lambda^2\approx0, \qquad p^\mu_\Lambda\approx0.
\end{equation}
Supposing that $\Lambda D\ne0$ (analog of the standard light-cone
singularity), one can extract two 2CC from Eq. (3):
$\Lambda^2\approx0$, $p_\Lambda D\approx0$ and nine 1CC: $\tilde
p^\mu_\Lambda\equiv p^\mu_\Lambda-\frac{p_\Lambda D}{\Lambda
D}\Lambda^\mu\approx0$ (there is identity $D_\mu\tilde
p^\mu_\Lambda\equiv0$). Thus, Eq. (3) provides a nondynamical character
of the auxiliary variables. Since by construction
$(\Lambda_\mu+D_\mu){\Gamma^\mu}_{\alpha\beta}$ is a nondegenerate
matrix, the constraints $L_\alpha\approx0$ are equivalent to
\begin{eqnarray}
&& L^{(1)\alpha}\equiv D_\mu\tilde\Gamma^{\mu\alpha\beta}L_\beta
\approx 0,\\
&& L^{(2)\alpha}\equiv \Lambda_\mu\tilde\Gamma^{\mu\alpha\beta}L_\beta
\approx 0,
\end{eqnarray}
where among 1CC $L^{(1)}\approx0$ and 2CC $L^{(2)}\approx0$ there are in
eight linearly independent. To supplement the IR1C in Eq. (4) to
irreducible, let us further introduce a pair of spinor variables
$(\chi^\alpha,p_{\chi\alpha})$ subject to constraints\footnote{After that,
instead of $\tilde p_\Lambda{}^\mu\approx0$, the following constraints:
\[ {p_\lambda}^\mu -\frac{p_\Lambda D}{\Lambda D}\Lambda^\mu +
\frac 1{2\Lambda D}p_\chi\tilde\Gamma^\nu D_\nu\chi\approx0 \]
will be first class.}
\begin{equation}
p_{\chi\alpha}\approx0, \qquad T_\alpha\equiv \Lambda_\mu
{\Gamma^\mu}_{\alpha\beta}\chi^\beta\approx0.
\end{equation}
These equations contain 8 independent 1CC among $p^{(1)}_\chi\equiv
\Lambda_\mu\tilde\Gamma^\mu p_\chi\approx0$ and 8+8 independent 2CC
among $p^{(2)}_\chi\equiv D_\mu\tilde\Gamma^\mu p_\chi\approx0$,
$\Lambda_\mu\Gamma^\mu\chi\approx0$. Note that the covariant gauge
$D_\mu\Gamma^\mu\chi=0$ may be imposed after that the full system
(constraints + gauge) is equivalent to $p_\chi\approx0$,
$\chi\approx0$.

Within the framework of the extended formulation it is possible to combine
part of the constraints into irreducible sets. Actually, taking into
account that the matrix $(\Lambda_\mu+D_\mu)\Gamma^{\mu\alpha\beta}$ is
nondegenerate one concludes that Eqs. (4) and (6) are equivalent to

\begin{eqnarray}
&& \Phi^\alpha\equiv L^{(1)\alpha}+p^{(1)\alpha}_\chi=D_\mu
\tilde\Gamma^{\mu\alpha\beta}L_\beta+\Lambda_\mu
\tilde\Gamma^{\mu\alpha\beta} p_{\chi\beta}\approx0,\\
&& G^\alpha\equiv L^{(2)\alpha}+p^{(2)\alpha}_\chi=\Lambda_\mu
\tilde\Gamma^{\mu\alpha\beta}L_\beta+D_\mu
\tilde\Gamma^{\mu\alpha\beta} p_{\chi\beta}\approx0,\\
&& T_\alpha\equiv \Lambda_\mu{\Gamma^\mu}_{\alpha\beta}\chi^\beta
\approx0,
\end{eqnarray}
where the $\Phi^\alpha\approx0$ ($G^\alpha\approx0$) are 16 irreducible 1CC
(2CC) and the $T_\alpha\approx0$ include 8 linearly independent 2CC. As a
result, for the modified formulation (3), (7)--(9) fermionic first and second
class constraints are splitted in manifestly covariant fashion, so that
1CC are irreducible. Note that the situation with 2CC does not became
``worse'' as compared to the initial formulation.
Some comments are in order.
\begin{enumerate}
\def\labelenumi{(\roman{enumi})}
\item For the case of CBS superparticle constraints without light-cone
singularities: $\Lambda^2\approx0$, $\Lambda D-1\approx 0$,
$p^\mu_\Lambda\approx0$ instead of Eq. (3) seem to be more suitable.
\item By making use of the vectors $D^\mu$, $\Lambda^\mu$ subject to
constraints from Eqs. (1) and (3), the true projectors
\begin{equation}
\begin{array}{l}
{\Pi^\pm}_\alpha{}^\beta=\frac 12\Big(1\pm \frac 1{2b}\Gamma^{\mu\nu}
D_\mu\Lambda_\nu\Big)_\alpha{}^\beta, \qquad b=\sqrt{D^2\Lambda^2
-(D\Lambda)^2},\\
1=\Pi^++\Pi^-, \qquad (\Pi^\pm)^2=\Pi^\pm, \qquad
\Pi^+\Pi^-=0,\end{array}
\end{equation}
can be constructed and applied in the previous scheme instead of the
matrices $D_\mu\Gamma^\mu$, $\Lambda_\mu\Gamma^\mu$.
\item For some concrete models the described trick may be realized
without introducing the variables $(\Lambda^\mu,p_{\Lambda\mu})$. For
the case of $D=9$ massive superparticle with Wess--Zumino term there
exists constant Lorentz-invariant matrix $z^{\alpha\beta}$ [4, 6]. It
can be used for constructing the covariant projectors and splitting the
constraints. For the case of GS superstring we may choose
$\Lambda^\mu\equiv p^\mu-\pa_1x^\mu$, since
$(p^\mu-\pa_1x^\mu)^2\approx0$ is one of the super Virasoro
constraints.
\end{enumerate}

Proceeding to the case of models with IR1C only, let us suppose
\begin{eqnarray}
& L^{(1)\alpha}\equiv D_\mu\tilde\Gamma^{\mu\alpha\beta}(p_\theta-
iB_\mu\Gamma^\mu\theta)_\beta\approx0, \qquad D^\mu D_\mu\approx0,\\
& \{L^{(1)\alpha},L^{(1)\beta}\}\approx0,
\end{eqnarray}
instead of Eqs. (1), (2) (for instance, Siegel superparticle
corresponds to the choice $D^\mu=B^\mu=p^\mu$). Repeating the procedure
described above,
we get the following constraints system in extended by the variables
$(\Lambda^\mu,p_{\Lambda\mu})$ and $(\chi^\alpha,p_{\chi\alpha})$
phase space:
\begin{eqnarray}
&&\begin{array}{l} D^2\approx0, \qquad \Lambda^2\approx0, \qquad
p^\mu_\Lambda\approx0,\\
\Phi^\alpha\equiv L^{(1)\alpha}+\Lambda_\mu\tilde\Gamma^{\mu\alpha\beta}
p_{\chi\beta} \approx0,\end{array}\\
&& p^{(2)\alpha}_\chi\equiv D_\mu\tilde\Gamma^{\mu\alpha\beta}
p_{\chi\beta}\approx0,\\
&& T_\alpha\equiv \Lambda_\mu\Gamma^\mu_{\alpha\beta}\chi^\beta
\approx0.
\end{eqnarray}
In contrast to the previous case, it is impossible to combine 2CC
from Eqs. (14), (15) in a manifestly covariant way because they belong
to different inequivalent representations of $SO(1{,}9)$ group of
opposite chirality. To avoid the problem, one needs further extension
of the phase space in order to construct a matrix for lowering (raising)
spinor indices. Let us introduce vectors $(C^\mu,p^\mu_C)$ subject
to constraints
\begin{equation}
C^2-1\approx0, \qquad C\Lambda\approx0, \qquad CD\approx0, \qquad
p^\mu_C\approx0.
\end{equation}
(Note that for all the above mentioned concrete models the condition
$\{D^2,C^\mu D_\mu\}\sim C^\mu D_\mu$ hold, and consequently
the constraint $D^2\approx0$ is first class as before.) The full system
of bosonic constraints (13) and (16) can be splitted into first and
second class sets
\begin{eqnarray}
&&\begin{array}{l} \Lambda^2\approx0, \quad p_\Lambda D\approx0; \qquad
C^2-1\approx0, \quad p_CC\approx0;\\
C\Lambda\approx0, \quad \displaystyle p_CD\approx0; \quad CD\approx0,
\quad p_C\Lambda\approx0;\end{array}\\[1ex]
&&\begin{array}{l} \tilde p^\mu_\Lambda\equiv p^\mu_\Lambda-\displaystyle
\frac{p_\Lambda D}{\Lambda D}\Lambda^\mu-(p_\Lambda C)C^\mu+
\frac 1{2\Lambda D}p_\chi\tilde\Gamma^{\nu}D_{\nu}\Gamma^{\mu}\chi\approx0,\\
\tilde p^\mu_C\equiv p^\mu_C-\displaystyle\frac{p_CD}{\Lambda D}\Lambda^\mu
-(p_CC)C^\mu-\frac{p_C\Lambda}{\Lambda D}D^\mu\approx0,\\
\displaystyle\frac{p_CD}{\Lambda D}-p_\Lambda C+\frac 1{2\Lambda D}p_\chi
C_{\nu}\tilde\Gamma^{\nu}\Gamma^{\mu}D_{\mu}\chi\approx0,\end{array}
\end{eqnarray}
where 1CC in Eq. (18) are first stage of reducibility. The reducibility
is described by equations $\Lambda\tilde p_\Lambda=C\tilde p_\Lambda=
\Lambda\tilde p_C=D\tilde p_C=C\tilde p_C=0$ which hold modulo 2CC (17).
Manifestly covariant quantization of the sector (as well as Eq. (3))
may be carried out along the lines of Ref. 10.

Having in our disposal the invertible matrix
$C_\mu{\Gamma^\mu}_{\alpha\beta}$, we pass from Eq. (14) to equivalent
constraints
\begin{equation}
p^{(2)}_{\chi\alpha}\equiv(C_\mu\Gamma^\mu\,D_\nu\tilde\Gamma^\nu
p_\chi)_\alpha\approx0,
\end{equation}
which may be covariantly combined now with Eq. (15). Thus, the resultant
formulation being equivalent to Eqs. (11), (12) takes the form
\begin{eqnarray}
&& \begin{array}{l} D^2=0, \qquad \Lambda^2=0,\qquad p^\mu_\Lambda=0,\\
C^2=1, \qquad C\Lambda=CD=0, \qquad p^\mu_C=0,\end{array}\\
&& \Phi^\alpha\equiv L^{(1)\alpha}+\Lambda_\mu\tilde\Gamma^{\mu\alpha\beta}
p_{\chi\beta}=0,\\
&& G_\alpha\equiv\Lambda_\mu{\Gamma^\mu}_{\alpha\beta}\chi^\beta+
(C_\mu\Gamma^\mu D_\nu\tilde\Gamma^\nu p_\chi)_\alpha=0,
\end{eqnarray}
where the Poisson bracket of the constraints (22) is
$\{G_\alpha,G_\beta\}=-2(\Lambda D)C_\mu{\Gamma^\mu}_{\alpha\beta}$. As
a result, the task of quantization of a model with infinitely reducible
fermionic
first class constraints has been reduced to quantization of splitted
(in manifestly covariant way) and irreducible first and second class
constraints (Eq. (21) and Eq. (22), respectively).

As an example of application the scheme proposed, we shall consider
the GS superstring, CBS superparticle and Siegel superparticle. For
each case, a modified action and its local symmetries in suitably
enlarged configuration space will be presented. By passing from the
Lagrangian formalism to the Hamiltonian one, we prove an equivalence of the
modified and initial formulations and then demonstrate an applicability
of the supplementation scheme.

\def\ve{\varepsilon}
\def\pa{\partial}
\def\La{\Lambda}
\def\de{\delta}
\subparagraph{{\boldmath$D=10,N=1$} Green--Schwarz superstring.}
Consider a covariant action of the form
\begin{equation}
\begin{array}{l} S=S_{\rm GS}+S_{\rm add}=\displaystyle\int d^2\sigma
\Big[-\frac 1{2\sqrt{-g}}g^{ab}\Pi^\mu_a\Pi_{b\mu}-i\ve^{ab}\pa_ax_\mu
(\theta\Gamma^\mu\pa_b\theta)-\\
\qquad\qquad-\displaystyle\frac 12 \La_\mu\ve^{ab}{F^\mu}_{ab}-
\phi\La^\mu\La_\mu\Big],\\
{\Pi_a}^\mu\equiv\pa_ax^\mu-i\theta\Gamma^\mu\pa_a\theta,\\
{F^\mu}_{ab}\equiv\pa_a{A_b}^\mu-\pa_b{A_a}^\mu-i\pa_a\theta\Gamma^\mu
\chi_b+i\pa_b\theta\Gamma^\mu\chi_a+i\chi_a\Gamma^\mu\chi_b,\end{array}
\end{equation}
where $S_{\rm GS}$ is the standard GS action [1]. It was chosen
$\ve^{01}=-1$, and the following auxiliary variables were introduced:
scalar $\phi$; $D=10$ vector $\La^\mu$; $D=2$ and $D=10$ vector
${A_a}^\mu$; $D=2$ vector and $D=10$ Majorana--Weyl spinor
${\chi_a}^\alpha$. As will be seen, the only essential variables for the
supplementation scheme are ${A_1}^\mu$ and ${\chi_1}^\alpha$,
all other prove to supply  $D=2$ reparametrization invariance of the action
(23). Global symmetries of the theory (23) are standard $D=10,N=1$
super Poincar\'e transformations.

Local bosonic symmetries are $D=2$ reparametrizations, Weyl symmetry, and the
following transformations with parameters $\xi^\mu(\sigma)$,
$\omega_a(\sigma)$:
\begin{equation}
\begin{array}{l} \de {A_a}^\mu=\pa_a\xi^\mu+\omega_a\La^\mu,\\
\de\phi=\displaystyle\frac 12 \ve^{ab}\pa_a\omega_b.\end{array}
\end{equation}
These symmetries are reducible because their combination with parameters
of a special form: $\omega_a=\pa_a\omega$, $\xi^\mu=-\omega\La^\mu$, is
a trivial symmetry: $\de_\omega{A_a}^\mu=-\omega\pa_a\La^\mu$,
$\de_\omega\phi=0$ (note that $\pa_a\La^\mu=0$ is one of the equations
of motion). Thus, Eq. (24) includes 11 essential parameters which
correspond to primary 1CC ${p^\mu}_{A_0}\approx0$, $p_\phi\approx0$
(see below). Besides, there are local symmetries with fermionic
parameters ${\kappa^-}_\alpha{}^a \equiv P^{-\,ab}\kappa_{\alpha
b}(\sigma)$ (where $P^{\pm\,ab}\equiv(g^{ab}/\sqrt{-g}\pm\ve^{ab})$) and
${S^+}_\alpha{}^a(\sigma)$:
\begin{eqnarray}
&& \begin{array}{l} \de\theta=\Pi_{\mu a}\tilde\Gamma^\mu\kappa^{-a},\\
\de x^\mu=i\theta\Gamma^\mu\de\theta,\\
\de\Big(\displaystyle\frac{g^{ab}}{\sqrt{-g}}\Big)=4iP^{-ac}
(\pa_c\theta\kappa^{-b}),\\
\de\chi_a=\pa_a(\de\theta)+\La_\mu\tilde\Gamma^\mu{\kappa^-}_a,\\
\de {A^\mu}_a=i\theta\Gamma^\mu\pa_a(\de\theta),\\
\de\phi=-i\epsilon^{ab}(\pa_a\theta-\chi_a){\kappa^-}_b;\end{array}\\[6pt]
&& \begin{array}{l} \de{\chi_a}^\alpha=\La_\mu\tilde\Gamma^{\mu\,
\alpha\beta}{S^+}_{\beta a},\\
\de\phi=-i\ve^{ab}(\pa_a\theta^\alpha-{\chi_a}^\alpha){S^+}_{\alpha b}.\end{array}
\end{eqnarray}

Equation (25) is generalization of Siegel $\kappa$-symmetry [20] to
the present case. In our formulation it is irreducible (with 16
essential parameters), and looks like a gauge symmetry (with the gauge
field to be $\chi_a$, as is seen from its transformation law). The
transformations (26) are reducible and incorporate only 8 essential
parameters among the ${S^+}_\alpha{}^a$, because rank $\La_\mu\Gamma^\mu=8$
on-shell, as a consequence of the equation of motion $\La^2=0$. Note that
it is not necessary to take care of the analog of the transformations (26)
with the parameters ${S^-}_\alpha{}^a$ since they have already been included
into Eq. (25) [18]. Thus, the presented transformations with 16+8 parameters
exhaust all the essential fermionic symmetries of the model, because
namely this number of primary fermionic 1CC will occur in the
Hamiltonian formalism.

By direct application of the Dirac--Bergmann algorithm [19] one gets
the Hamiltonian
\begin{eqnarray}
H&=&-\displaystyle\frac 1{g^{00}}\Big(\frac{\sqrt{-g}}2
(\hat p^2+\Pi^2_1)-g^{01}\hat p_\mu\Pi_1^\mu\Big)-\pa_1\La_\mu
A_0^\mu-\cr
&-& i\La_\mu(\pa_1\theta-\chi_1)\Gamma^\mu\chi_0+\phi\La^2+
\lambda^{ab}(p_g)_{ab}+\lambda_0p_\phi+\lambda_{1\mu}{p_\La}^\mu+\cr
&+&\lambda_{2\mu}{p^\mu}_{A_0}+\lambda_{3\mu}({p^\mu}_{A_1}-\La^\mu)+
p_{\chi0}\sigma_0+p_{\chi1}\sigma_1+L\sigma_3,
\end{eqnarray}
where $\lambda,\sigma$ are Lagrange multipliers for primary
constraints, and it was denoted  $\hat p^\mu\equiv
p^\mu-i\theta\Gamma^\mu\pa_1\theta$. The full set of constraints can be
written in the form
\begin{eqnarray*}
&& (p_g)_{ab}=0, \qquad p_\phi=0, \qquad {p^\mu}_{A_0}=0, \qquad
p_{\chi0\alpha}=0;\qquad\qquad\qquad\qquad\qquad\quad (28.a)\\
&& ({p^\mu}_{A_1})^2=0, \qquad \pa_1{p^\mu}_{A_1}=0;
\qquad\qquad\qquad\qquad\qquad\qquad\qquad\qquad\qquad\qquad\quad (28.b)\\
&& {p_\La}^\mu=0, \qquad {p^\mu}_{A_1}-\La^\mu=0;
\qquad\qquad\qquad\qquad\qquad\qquad\qquad\qquad\qquad\qquad\quad (28.c)\\
&& H_1\equiv(\hat p^\mu+\Pi_1^\mu)^2-4L_\alpha\pa_1\theta_\alpha=0,
\qquad H_2\equiv(\hat p^\mu-\Pi_1^\mu)^2=0;\qquad\qquad\qquad\quad (28.d)\\
&& L_\alpha\equiv p_{\theta\alpha}-\pa_1p_{\chi1\alpha}-i(p^\mu+\Pi_1^\mu)
(\theta\Gamma^\mu)_\alpha+ip_{A_1\mu}(\pa_1\theta\Gamma^\mu)_\alpha=0;
\qquad\qquad\qquad\quad (28.e)\\
&& p_{\chi_1\alpha}=0, \qquad p_{A_1\mu}(\pa_1\theta-\chi_1)^\beta
{\Gamma^\mu}_{\beta\alpha}=0, \qquad\qquad\qquad\qquad\qquad\qquad\qquad\qquad (28.f)
\end{eqnarray*}
\stepcounter{equation}
where some of the initial constraints were exchanged on equivalent ones
to simplify the Poisson brackets algebra. There are 1CC in Eqs. (28.a),
(28.b), (28.d) and a trivial pair of 2CC in Eq. (28.c). Among 11
equations (28.b) only 10 are functionally independent, due to the identity
$\pa_1[p^2_{A_1}]-2p^\mu_{A_1}[\pa_1p_{A_1\mu}]\equiv0$. Poisson
brackets of the constraints (28.d), (28.e) are identical to those the
GS superstring [1, 5], in particular, $\{L_\alpha,L_\beta\}=2i(\hat
p^\mu+\Pi^\mu_1) {\Gamma^\mu}_{\alpha\beta} \de(\sigma-\sigma')$. At
last, the constraints (28.f) can be represented in equivalent form
\begin{eqnarray}
&& p_{A_1\mu}\tilde\Gamma^\mu p_{\chi1}=0;\\[6pt]
&& \begin{array}{l} (\hat p^\mu+\Pi^\mu_1)\tilde\Gamma^\mu p_{\chi1}=0;\\
p_{A_1\mu}(\pa_1\theta-\chi_1)\Gamma^\mu=0,\end{array}
\end{eqnarray}
with 8 independent 1CC among Eqs. (29) and 8+8 independent 2CC among
Eqs. (30).

To investigate dynamics of the theory, we pass to light-cone
coordinates $\big(x^\mu\to(x^+,x^-,x^i)$, $i=1,\dots,8$,
$\theta^\alpha\to (\theta_a,\bar\theta_{\dot a})$, $a,\dot
a=1,\dots,8\big)$, write out equations of motion for all variables
with the help of Eq. (27), take into account the full constraints
system (28), and impose gauge fixing conditions to 1CC. A
self-consistent gauge choice is
\begin{equation}
\begin{array} {l}
g^{\alpha\beta}=\eta^{\alpha\beta}, \qquad \phi=1/2, \qquad
A^\mu_0=0, \qquad \chi^\alpha_0=0,\\
\theta_a=0, \qquad \pa_1\bar\theta_{\dot a}-\bar\chi_{1\dot a}=0,\\
A^-_1=\tau, \qquad A^+_1=A^i_1=0,\\
x^+=-P^+\tau, \qquad p^+=P^+={\rm const}\ne 0.\end{array}
\end{equation}

After tedious calculations [18] one gets exactly the GS superstring
dynamics for $x^i,p^i,\bar\theta_{\dot a}$ variables
\begin{equation}
\pa_0x^i=-p^i, \qquad \pa_0p^i=-\pa_1\pa_1x^i, \qquad
(\pa_0+\pa_1)\bar\theta_{\dot a}=0,
\end{equation}
while all other variables and Lagrange multipliers are expressed
through them by means of algebraic equations. Note that in the gauge
chosen, the relations $p^-_{A_1}=1$, $p_{A_1}{}^+=p_{A_1}{}^i=0$, $\hat
p^++\Pi^+_1=P^+\ne0$, or, in covariant form
\begin{equation}
p_{A_1\mu}(\hat p^\mu+\Pi_1^\mu)\ne0
\end{equation}
hold. Return now to constraints (28.e), (29), (30), and
rewrite them in equivalent form with the use of Eq. (33):
\begin{eqnarray}
&& (\hat p_\mu+\Pi_{1\mu})\tilde\Gamma^\mu L+p_{A_1\mu}\tilde\Gamma^\mu
p_{\chi1}=0,\\ [1ex]
&& \begin{array}{l} p_{A_1\mu}\tilde\Gamma^\mu L+(\hat
p_\mu+\Pi_{1\mu})\tilde\Gamma^\mu p_{\chi1}=0,\\
p_{A_1\mu}(\pa_1\theta-\chi_1)\Gamma^\mu=0.\end{array}
\end{eqnarray}
As a result, for the modified formulation of the GS superstring (23)
fermionic constraints are splitted into first and second class (Eqs.
(34) and (35), respectively) in a manifestly covariant way, so that the
1CC are irreducible.

An interesting peculiarity of the presented formulation is that it
possesses 16+8 fermionic reducible symmetries (25), (26), while first
class constraints in the Hamiltonian formalism turn out to be irreducible.
The reason is that except
16+8 primary 1CC $p_{\chi_0}=0$, $\La_\mu\tilde\Gamma^\mu
p_{\chi_1}=0$, corresponding to the symmetries, there appear 8
secondary 1CC $(\hat p_\mu+\Pi_{1\mu})\tilde\Gamma^\mu L=0$. The two
reducible sets are simply combined into irreducible one in the
resultant system (34). This situation is opposite to the case of
Siegel superparticle in the initial formulation [12, 13] for that
symmetries are irreducible, while in the Hamiltonian formalism there arise
reducible secondary first class constraints.

\subparagraph{{\boldmath$D=10,N=1$} Casalbuoni--Brink--Schwarz superparticle.}

In this case we consider the following action:
\begin{equation}
S=\int d\tau\Big[-\frac 1{2e}\Pi^\mu\Pi_\mu-\omega-i\La_\mu\chi_1
\Gamma^\mu(\chi_0+\dot\theta)-\phi\La^2\Big],
\end{equation}
where $\Pi^\mu\equiv\dot x^\mu-i\theta\Gamma^\mu\dot\theta+\omega
\La^\mu$. The variables $\La^\mu$ and ${\chi_1}^\alpha$ turn out to be
essential for realization of the supplementation scheme, while the
variables $\omega$, $\phi$ ${\chi_0}^\alpha$ are in fact Lagrange
multipliers which will supply appearance of the necessary constraints (3),
(6). Similarly to previous case, there are reducible fermionic
symmetries with 8+8+8 intrinsic parameters
\begin{eqnarray}
&& \begin{array}{ll} \de\theta^\alpha=\Pi_\mu\tilde\Gamma^{\mu\,
\alpha\beta}\kappa_\beta, & \de x^\mu=i\theta\Gamma^\mu\de\theta,\\
\de e =-2ie\dot\theta\kappa, & \de\chi_0=-(\de\theta)^{\cdot};\end{array}\\[1ex]
&& \begin{array}{ll} \de{\chi_0}^\alpha=\La_\mu\tilde\Gamma^{\mu\,
\alpha\beta}S_{0\beta}, & \de\phi=i{\chi_1}^\alpha S_{0\alpha};\\
\de{\chi_1}^\alpha=\La_\mu\tilde\Gamma^{\mu\,\alpha\beta}S_{1\beta}, &
\de\phi=iS_{1\alpha}(\chi_0+\dot\theta)^\alpha;\end{array}
\end{eqnarray}
that exactly correspond to the independent primary first class
constraints of the model (see below).

Remarkably, the action (36) leads only to the desired constraints (3),
(6) for the variables $\La^\mu$ and ${\chi_1}^\alpha$ and does not
imply any other constraints as it could be expected. We give here a
detailed discussion
of the Dirac--Bergmann algorithm for the model because some accuracy is
necessary in treating its constraints system. The canonical Hamiltonian is
\begin{eqnarray}
&& H=-\displaystyle\frac e2 p^2-\omega(p\La-1)+i\La_\mu\chi_1
\Gamma^\mu\chi_0+\phi\La^2+\xi_0p_e+\xi_1p_\omega +\xi_2p_\phi+\cr
&&\qquad +\xi_{3\mu}{p_\La}^\mu+p_{\chi0}\sigma_0+p_{\chi1}\sigma_1+
L_\alpha{\sigma_3}^\alpha,
\end{eqnarray}
where $\xi_i$, $\sigma_i$ are Lagrange multipliers corresponding to the
primary constraints, and we denoted $L\equiv p_\theta-ip_\mu\theta\Gamma^\mu
-i\La_\mu\chi_1\Gamma^\mu\approx0$. From requirement of preservation in
time of the primary constraints, we get the secondary ones
\begin{eqnarray}
& p^2=0, \qquad \La^2=0, \qquad p\La-1=0,\\
& \La_\mu(\chi_1\Gamma^\mu)_\alpha=0,
\end{eqnarray}
and equations containing the Lagrange multipliers which can be represented
in the form
\begin{equation}
\begin{array}{l} \La_\mu\Gamma^\mu\sigma_1=0, \qquad \La_\mu\Gamma^\mu
(\chi_0-\sigma_3)=0,\\
p_\mu\Gamma^\mu\sigma_3=0, \qquad -i\chi_1\Gamma^\mu(\chi_0-\sigma_3)+
\omega p^\mu-2\phi\La^\mu=0.\end{array}
\end{equation}
(Multiplying the last equation by the $\La_\mu$ we get as a consequence
$\omega=0$ and the corresponding term may be omitted.) To analyze the
system, note first that by virtue of Eq. (40), the following
decompositions are possible:
\begin{equation}
\chi_0=\chi+\tilde\chi, \qquad \sigma_3=\sigma+\tilde\sigma,
\end{equation}
where the corresponding components obey
\begin{equation}
\La_\mu\Gamma^\mu\chi=p_\mu\Gamma^\mu\tilde\chi=\La_\mu\Gamma^\mu\sigma
=p_\mu\Gamma^\mu\tilde\sigma.
\end{equation}
Then, one can verify that Eqs. (42) are equivalent to
\begin{eqnarray}
& \tilde\sigma=\tilde\chi, \qquad \sigma=0, \qquad \omega=0,\\
& -i\chi_1\Gamma^\mu\chi-2\phi\La^\mu=0.
\end{eqnarray}
Moreover, passing to the light-cone coordinates and using $SO(8)$
notations for spinors it is easy to check that there is only one
independent equation in Eqs. (46), if the conditions
$\La_\mu\chi_1\Gamma^\mu=0$, $\La_\mu\Gamma^\mu\chi=0$, $\La^2=0$
hold, namely: $-i\sqrt{2}\bar\chi_{1\dot a}\bar\chi_{\dot a}-
2\phi\La^-=0$.
(Note that from the Eq. (40) it follows $\La^+\ne0$ or $\La^-\ne0$. For
definiteness we choose the latter.)

The derived constraint forms a pair of second class constraints with
$p_\phi=0$, and may be extracted from Eq. (42) in covariant way
as follows:
$-p_\mu\chi_1\Gamma^\mu\chi_0-2\phi=0$. Finally, a full constraints
system acquires the form
\begin{eqnarray*}
&& p_e=p_\omega=\omega=0, \qquad p_{\chi0}=0;
\qquad\qquad\qquad\qquad\qquad\qquad\qquad\qquad\qquad\qquad\quad (47.a)\\
&& p^2=0, \qquad p\La-1=0, \qquad \La^2=0, \qquad {p_\La}^\mu=0;
\qquad\qquad\qquad\qquad\qquad\qquad\quad (47.b)\\
&& L\equiv p_\theta-ip_\mu\theta\Gamma^\mu=0, \qquad p_{\chi1}=0, \qquad
\La_\mu\chi_1\Gamma^\mu=0; \qquad\qquad\qquad\qquad\qquad\qquad (47.c)\\
&& p_\phi=0, \qquad \Phi\equiv\phi+\displaystyle\frac i2 p_\mu\chi_1
\Gamma^\mu\chi_0=0.
\qquad\qquad\qquad\qquad\qquad\qquad\qquad\qquad\qquad (47.d)
\end{eqnarray*}
\stepcounter{equation}
The second constraint from Eq. (47.d) has nontrivial Poisson brackets
with some of the constraints (47.a), (47.c). To avoid the obstacle, let
us pass to the Dirac bracket associated with the pair (47.d)
\begin{equation}
\{A,B\}_D=\{A,B\}-\{A,p_\phi\}\{\Phi,B\}+\{A,\Phi\}\{p_\phi,B\}.
\end{equation}
Then, it is admissible to consider the constraints as strong equalities
and to resolve them. After that the variables $p_\phi,\phi$ can be
dropped. As is seen from Eq. (48), the Dirac brackets for the remaining
variables coincide with the Poisson ones.

As a result, for the model (36) we have got the desired constraints
system (47.a--c). Actually, Eq. (47.c) can be rewritten in equivalent
form
\begin{eqnarray*}
&& p_\mu\tilde\Gamma^\mu L+\La_\mu\tilde\Gamma^\mu p_{\chi1}=0;
\qquad\qquad\qquad\qquad\qquad\qquad\qquad\qquad\qquad\qquad\qquad\qquad (49.a)\\
&& \La_\mu\tilde\Gamma^\mu L+p_\mu\tilde\Gamma^\mu p_{\chi1}=0, \quad
\La_\mu\chi_1\Gamma^\mu=0, \qquad\qquad\qquad\qquad\qquad\qquad\qquad\qquad\qquad (49.b)
\end{eqnarray*}
\stepcounter{equation}
with the 1CC (49.a) being irreducible.

Dynamics of the model may be analyzed along the same lines as has been
done for the GS superstring above. Repeating all the needed steps, one
can verify that physical sector and corresponding equations of motion
of the modified formulation (36) are exactly the same as those of CBS
superparticle.

\subparagraph{Siegel superparticle.} Within the framework of developed
supplementation scheme, this is the most interesting example since a
reducible 2CC are absent in the initial formulation [12, 13]. As has
been shown before, constraints system of the enlarged formulation for
this case will include irreducible first and second class constraints
only. The action, which reproduces the desired constraints system looks
as follows:
\begin{eqnarray}
\lefteqn{S=\int d\tau\Big[-\displaystyle\frac 1{2e} \Pi^\mu\Pi_\mu+
i\dot\theta_R\theta_L-i\La_\mu\chi_1\Gamma^\mu(\chi_0+\dot\theta_R)-}\cr
&& -\omega_0-\phi\La^2-\omega_2(C^2-1)-\omega_3(C\La)\Big],
\end{eqnarray}
where $\Pi^\mu\equiv\dot x^\mu-i\theta_R\Gamma^\mu\dot\theta_R+i\psi_L
\tilde\Gamma^\mu\theta_L+\omega_0\La^\mu+\omega_1C^\mu$, and
$\theta_R^\alpha$, $\theta_{L\alpha}$ are Majorana--Weyl spinors of
opposite chirality. The theory is invariant under the following
generalization of Siegel $k$-symmetry:
\begin{eqnarray}
&& \de\theta_R^\alpha=\frac 1e \Pi_\mu(\tilde\Gamma^\mu\kappa_L)^\alpha,
\qquad \de\theta_{L\alpha}=-\frac 2{e^2}\kappa_{L\alpha}\Pi^2,\cr
&& \de X^\mu=i\theta_L\tilde\Gamma^\mu\kappa_L+i\theta_R\Gamma^\mu\de
\theta_R,\\
&& \de e=-\frac{4i}e \Pi_\mu\psi_L\tilde\Gamma^\mu\kappa_L, \qquad
\de\psi_{L\alpha}=\dot\kappa_{L\alpha}, \qquad
\de{\chi_0}^\alpha=-\de{\theta_R}^\alpha,\nonumber
\end{eqnarray}
that is irreducible as in the initial formulation [12, 13] as well as
under a pair of reducible symmetries analogous to Eq. (38) of the
previous case
\begin{equation}
\begin{array}{ll} \de{\chi_0}^\alpha=\La_\mu\tilde\Gamma^{\mu
\alpha\beta}S_{0\beta}, & \de\phi=i{\chi_1}^\alpha S_{0\alpha};\\
\de{\chi_1}^\alpha=\La_\mu S_{1\beta}\tilde\Gamma^{\mu \alpha\beta}, &
\de\phi=iS_{1\alpha}(\chi_1+\dot\theta_R)^\alpha;\end{array}
\end{equation}

Canonical analysis for the model turns out to be very similar to that of
the previous case. So, we present and discuss only the final answer for the
full system of constraints
\begin{eqnarray*}
&& p_{\omega i}=0, \qquad \omega_i=0, \qquad i=0,1,2,3;
\qquad\qquad\qquad\qquad\qquad\qquad\qquad\qquad\qquad (53.a)\\
&& p_\phi=0, \qquad \phi+\frac i2 p_\mu\chi_1\Gamma^\mu(\chi_0-p_\nu
\tilde\Gamma^\nu\psi_L)=0; \qquad\qquad\qquad\qquad\qquad\qquad\qquad\quad (53.b)\\
&& p_{\chi0\alpha}=0, \qquad {p_\psi}^\alpha=0, \qquad \La_\mu\Gamma^\mu
(\chi_0-p_\nu\tilde\Gamma^\nu\psi_L)=0; \qquad\qquad\qquad\qquad\qquad\qquad (53.c)\\
&& {p_L}^\alpha=0, p_{R\alpha}-ip_\mu(\theta_R\Gamma^\mu)_\alpha-
i\theta_{L\alpha}=0;
\qquad\qquad\qquad\qquad\qquad\qquad\qquad\qquad\qquad (53.d)\\[1ex]
&& \begin{array}{l} p_e=0, \qquad {p_\La}^\mu=0, \qquad pC^\mu=0,
\qquad \La^2=0,\\
p\La-1=0, \qquad C^2-1=0, \qquad pC=0, \qquad C\La=0;\end{array}
\qquad\qquad\qquad\qquad\qquad (53.e)\\[1ex]
&& p_\mu(p_R\tilde\Gamma^\mu)^\alpha=0, \qquad p_{\chi1\alpha}=0,
\qquad \La_\mu(\chi_1\Gamma^\mu)_\alpha=0.
\qquad\qquad\qquad\qquad\qquad\qquad\quad (53.f)
\end{eqnarray*}
\stepcounter{equation}
Here, we have a set of trivial 2CC (53.a); a pair of 2CC (53.d) which
is the same as for the Siegel superparticle in the initial formulation [12,
13], and a pair of 2CC (53.b) that may be treated similarly to Eq.
(47.d). Constraints system (53.c) consists of 24 independent 1CC and 16
independent 2CC which may be covariantly separated as follows:
\begin{eqnarray*}
&& p_\mu\Gamma^\mu p_\psi=0, \qquad \La_\mu\tilde\Gamma^\mu p_{\chi0}=0,
\qquad \La_\mu\Gamma^\mu(p_\nu\tilde\Gamma^\nu p_{\chi0}+p_\psi)=0,
\qquad\qquad\qquad\qquad (54.a)\\
&& \La_\mu\Gamma^\mu(p_\nu\tilde\Gamma^\nu p_{\chi0}-p_\psi)=0, \qquad
\La_\mu\Gamma^\mu(\chi0-p_\mu\tilde\Gamma^\mu\psi_L)=0.
\qquad\qquad\qquad\qquad\qquad\quad (54.b)
\end{eqnarray*}
\stepcounter{equation}
Choosing now gauge fixing conditions to 1CC (54.a) in the form
\begin{equation}
\La_\mu\tilde\Gamma^\mu\psi_L=0, \qquad p_\mu\Gamma^\mu\chi_0=0,
\qquad \La_\mu\Gamma^\mu(\chi_0+p_\nu\tilde\Gamma^\nu\psi_L)=0,
\end{equation}
one can check that the full system of equations (54.a), (54.b), and
(55) is equivalent to $p_\psi=0$, $\psi_L=0$, $p_{\chi0}=0$, $\chi_0=0$.
Assuming the gauge has been imposed, let us pass to the Dirac bracket
associated
with the constraints (53.a--c). Then the corresponding variables may be
neglected, while the Dirac brackets for the remaining variables exactly
coincide with the Poisson ones. The remaining constraints allow us to
realize the supplementation scheme for reducible constraints
$p_\mu(p_R\tilde\Gamma^\mu)=0$ which has been presented in the initial
formulation. Actually, by virtue of Eqs. (53.e) one can combine
fermionic constraints (53.f) into irreducible sets as follows:
\begin{equation}
\begin{array}{l} \Phi^\alpha\equiv p_\mu(p_R\tilde\Gamma^\mu)^\alpha+
\La_\mu(\tilde\Gamma^\mu p_{\chi1})^\alpha=0,\\
G_\alpha\equiv \La_\mu(\Gamma^\mu\chi_1)^\alpha+
C_\mu p_\nu(\Gamma^\mu\tilde\Gamma^\nu p_{\chi1})_\alpha=0.\end{array}
\end{equation}
Here we have 16 irreducible 1CC $\Phi^\alpha=0$ and 16 irreducible 2CC
$G_\alpha=0$ with the bracket $\{G_\alpha,G_\beta\}=-2C_\mu
\Gamma^\mu_{\alpha\beta}$.

Note in conclusion that for the case of the Lorentz group $SO(1,8)$ there
exists only one inequivalent spinor representation of minimal dimension
(the above mentioned matrix $z^{\alpha\beta}$ may be used for lowering
and raising spinor indices). It means, in particular, that for $D=9$
Siegel superparticle it is not necessary to introduce variables
$\La^\mu$, $C^\mu$. This fact may extremely simplify a task of quantum
realization of the scheme suggested.

\end{document}